# Anomalous Nernst Effects of [CoSiB/Pt] Multilayer Films

O. Kelekci[1]†¶, H. N. Lee[2]†, T. W. Kim[2], and H. Noh[1*]

[1]*Department of Physics and Graphene Research Institute, Sejong University, Seoul 143-747, Korea*

[2]*Department of Advanced Materials Science and Engineering, Sejong University, Seoul 143-747, Korea*

We report a measurement for the anomalous Nernst effects induced by a temperature gradient in [CoSiB/Pt] multilayer films with perpendicular magnetic anisotropy. The Nernst voltage shows a characteristic hysteresis which reflects the magnetization of the film as in the case of the anomalous Hall effects. With a local heating geometry, we also measure the dependence of the anomalous Nernst voltage on the distance $d$ from the heating element. It is roughly proportional to $1/d^{1.3}$, which can be conjectured from the expected temperature gradient along the sample from the heat equation.

† Equally contributed to the work.

¶ Present address: Faculty of Engineering and Architecture, Siirt University, Siirt 56100, Turkey

# 1. Introduction

The Hall effects in a ferromagnetic material display a unique characteristic that results from a peculiar response of the carrier spins in the presence of the spin-orbit coupling. In addition to the ordinary contribution which is linear in the applied magnetic field, there exists an anomalous contribution which is linear in the magnetization of the material and nonlinear in the applied magnetic field. There have been extensive studies on this anomalous Hall effects (AHE) [1], and more recently on its counterpart in a non-magnetic system, the spin Hall effects [2-7] as well. In those effects, a longitudinal electric field is driving the carriers through the sample system, and a spin-dependent transverse electric field arises as a result. The net transverse electric field is non-zero in a ferromagnetic system that has a spontaneous spin polarization, whereas, it is zero in a non-magnetic system in which the transverse electric field for each spin component cancels. On the other hand, a temperature gradient can also drive the carriers through the system and similar spin-dependent transverse electric fields can develop, which is known as the anomalous Nernst effects (ANE) in a ferromagnetic system and the spin Nernst effects in a non-magnetic system. Studies on these effects are relatively rare, but could provide some important information on the nature of the spin transport as combined with the heat transport in the emerging field of spin-caloritronics [8].

There has been few studies on the ANE in single crystals of spinel ferromagnets such as $CuCr_2Se_{4-x}Br_x$ [9], $Nd_2Mo_2O_7$, and $Sm_2Mo_2O_7$ [10], perovskites such as $La_{1-x}Sr_xCoO_3$ and $SrRuO_3$ [11], a diluted magnetic semiconductor $Ga_{1-x}Mn_xAs$ [12], and more recently, in a thin film of perpendicularly magnetized FePt [13]. However, experiments on a wider variety of materials which are important for applications such as magnetic recording devices are still lacking. In addition, previous experiments on the ANE were done by heating the one edge of a macroscopically sized sample which results in a constant temperature gradient and an averaged response of the system. A local heating of the sample could provide additional information on the thermal transport characteristics as recently reported by Weiler *et al.* [14], where a scanning laser beam was employed on a metallic and an insulating ferromagnetic thin films to map out local responses of the ANE and the inverse spin Hall effects.

In this paper, we present an experimental study on the anomalous Nernst effects in a perpendicular magnetic [CoSiB/Pt] multilayer film by employing local heating. With the epitaxial FePt film considered as a possible candidate for future perpendicular magnetic recording applications, the amorphous multilayer films such as [CoSiB/Pt] are of comparable significances due to their merits of having less grain boundaries and less interface roughness [15,16], thus, resulting in a better switching performance when used as a free layer in a magnetic tunnel junction (MTJ) device [17]. In our experiment, a lithographically defined heater at one end of the Hall-bar shaped [CoSiB/Pt] film was used to produce a local heating and the resulting Nernst voltage ($V_N$) was measured in the presence of a magnetic field ($B$) perpendicular to the film. The magnitude of $V_N$ is proportional to the power generated by the heater confirming the thermal origin, and it nonlinearly depends on the magnetic field showing a hysteresis which is coincident with the anomalous Hall voltage. The local heating geometry produces a non-constant temperature gradient and enables us to detect differences in the local signals of $V_N$ as a function of the distance from the heater. The magnitude of $V_N$ is roughly proportional to $1/d^{1.3}$. This behavior suggests that the heat transport is in a crossover between the 2D and the 3D case as conjectured from the solutions to the heat equation with a point heat source.

## 2. Experiment

The ferromagnetic multilayer film of [CoSiB/Pt] was deposited on an oxidized Si wafer with a DC magnetron sputtering method. After the deposition of buffer layers with 50 Å of Ta and 30 Å of Pt, the repeated layers of 6 Å-thick CoSiB and 14 Å-thick Pt were deposited. A detailed, perpendicular magnetic property of these multilayer films have been recently reported [18]. In our measurements, samples with two different numbers ($N$) of the [CoSiB/Pt] layers were used, $N$=7 and 15. The multilayer film was fabricated into a Hall bar-shaped device with photolithography and ion-milling. Then, a separate tungsten heater pattern was made across the longitudinal arm of the Hall bar at one end. Figure 1 (a) shows the schematic measurement setup of the Nernst effect, and Figure 1 (b) shows the optical microscope image on the active region of the device. The width of the Hall bar is 10 μm and the distance between the adjacent voltage contacts is 20 μm. The nearest voltage contacts from the heater are 60 μm away. By applying a current through the heater and a magnetic field perpendicular to the film, we measured the Nernst voltages from the lateral contacts. All the measurements were performed at room temperature using a Keithley 2400 source meter and an electromagnet.

## 3. Results and Discussion

In Fig. 1 (c) and (d), we show $V_N$ as a function of $B$ measured with a heater current ($I_h$) of 30 mA for $N$=7 and 15, respectively, and the Hall voltage ($V_H$) measured with a drive current of 50 μA as well. There is a clear hysteresis of $V_N$ reflecting the anomalous contribution that results from the magnetization of the film, as in the case of $V_H$. We note that the drive current flows along the Hall bar when $V_H$ is measured but there is no drive current when $V_N$ is measured. Instead, the temperature gradient induced by the heater makes a thermal current flow and the electronic contributions of the thermal current along the Hall bar result in the developments of $V_N$ in the presence of $B$.

To make sure that the occurring $V_N$ signal is due to the thermal current, we varied the $I_h$ and measured the change in $V_N$. In Fig. 2 (a), we show $V_N$ as a function of $B$ measured with different values of $I_h$. The saturated value of $V_N$ increases with increasing $I_h$, but the dependence is nonlinear (Fig. 2(b)). In fact, the magnitude of $V_N$ depends on the power generated by the heater as shown in Fig. 2 (c), in which the slope of the best linear fit in a log-log scale is about 1.15. This confirms that the $V_N$ signal measured is from the thermal origin. We note that the coercive field ($H_c$) extracted from the data in Fig. 2 (a) also depends on the $I_h$. As shown in Fig. 3, $H_c$ is higher and close to the value obtained from the anomalous Hall measurements when $I_h$ is small, and becomes lower when $I_h$ increases. This is believed to result from the raised temperature at the position of probing contacts when a larger heater current is used. The exact temperature at each probing contact is difficult to measure in this micrometer-sized sample, which requires a complicated design of the sample incorporating on-chip thermometers. We suggest a possibility that the signal of $V_N$ itself could provide a temperature measurement in comparison with the temperature dependence of $H_c$ determined from other independent measurements such as the magnetization or the anomalous Hall effects.

Even though we cannot measure the temperature at the position of each probing contact, the dependence of the temperature gradient on the distance from the heater can be obtained by measuring $V_N$ from different probing contacts since $V_N$ is proportional to the temperature gradient. When we measure $V_N$ from other pairs of contacts which are at different distances from the heater, the results are shown in Fig. 4 (a). We observed similar anomalous behaviors of $V_N$ no matter how far the contacts are away from the heater, but the magnitude of $V_N$ decreases with distance. This implies that the temperature gradient is not constant along the Hall bar direction. It appears a bit odd that the temperature gradient is not constant if one considers a simple one dimensional heat conduction problem. It may be expected that the temperature profile along the longitudinal direction of the Hall bar would be linear as a function of the distance from the heater and the temperature gradient is constant. We note that the heat conduction in our sample occurs not only along the Hall bar direction but also to all directions on the surface and inside of the sample resulting in a temperature gradient which depends on the position. In most of the experiments on the anomalous Nernst effect, one side of the large sample was heated and the resulting temperature gradient was assumed to be a constant of $\Delta T / \Delta x$, where $\Delta T$ is the temperature difference between the heated and the cold ends, and $\Delta x$ is the distance between the two ends. In such an experiment, a detailed spatial variation of $V_N$ cannot be observed. The use of local heater in our experiment enables us to measure the changes of $V_N$ with position.

To see the detailed dependence of $V_N$ on the distance from the heater, we plot the saturated values of $V_N$ as a function of the distance (*d*) from the heater in Fig. 4 (b) and (c) in a linear and a log-log scale, respectively. From the slope in Fig. 4 (c), we deduce that $V_N$ is approximately proportional to $1/d^{1.3}$. To understand this behavior, we consider the heat equation. The heat equation that describes a heat conduction problem is a partial differential equation for temperature (*T*) with respect to time and position, and is given by

$$\nabla^2 T + \frac{\dot{q}}{k} = \frac{\rho c_p}{k} \frac{\partial T}{\partial t} \quad ,$$

where $k$ is the thermal conductivity, $\rho$ the density, $c_p$ the heat capacity, and $\dot{q}$ is the energy generation rate (or power) of a heat source. In the steady state, the equation becomes

$$\nabla^2 T + \frac{\dot{q}}{k} = 0 \quad,$$

which is the Poisson equation. Since it is impossible to find the exact solution to this equation that is appropriate to our sample geometry and the boundary, we make an assumption that the heater is point-like and the size of the sample is infinite to find a solution and approximately contrast that with our result. The fundamental solution to the equation in an unbounded space with a point heat source of $\dot{q} = P_0 \delta(\vec{x})$ is given by $T = (P_0/k)(1/4\pi d)$ in 3D and $T = -(P_0/k)(1/2\pi)\ln d$ in 2D. The temperature gradient is therefore proportional to $1/d^2$ in 3D and $1/d$ in 2D. Our results correspond to somewhere in-between the 2D and the 3D case. The overall planar geometry of the sample including the substrates makes it adequate to consider the 2D case, but the non-negligible heat conduction along the perpendicular direction appears to add a correction towards the 3D case as well.

## 4. Conclusion

We have measured the anomalous Nernst effects in [CoSiB/Pt] multilayer films and clearly observed the anomalous contributions that are proportional to the perpendicular magnetization. The magnitude of the Nernst voltages, being proportional to the power of the local heater, also showed a dependence on the position relative to the heater, which implies a non-constant temperature gradient induced in the sample. The detailed dependence on the distance from the local heater suggested that the heat conduction in the sample has a character in-between the 2D and the 3D case. Finally, we did not systematically address the dependence on the number of layers since only two different values of $N$ were used in this study, and the micrometer-sized samples of Hall bar geometry made it difficult to measure the exact temperatures at each probing contact, which is a crucial step for systematic comparisons between different samples under the same conditions. Further studies in such a direction will be necessary in the future.

## Acknowledgements

This work was supported by Basic Science Research Program (2010-0025340) and Priority Research Centers Program (2010-0020207) through the National Research Foundation of Korea (NRF) funded by the Ministry of Education, Science and Technology.

## References

[1] N. Nagaosa, J. Sinova, S. Onoda, A. H. MacDonald, and N. P. Ong, Rev. Mod. Phys. **82**, 1539 (2010).

[2] M. I. Dyakonov and V. I. Perel, Phys. Lett. **35A**, 459 (1971).

[3] J. E. Hirsch, Phys. Rev. Lett. **83**, 1834 (1999).

[4] Y. K. Kato, R. C. Myers, A. C. Gossard, and D. D. Awschalom, Science **306**, 1910 (2004).

[5] J. Wunderlich, B. Kaestner, J. Sinova, and T. Jungwirth, Phys. Rev. Lett. **94**, 047204 (2005).

[6] S. O. Valenzuela and M. Tinkham, Nature **442**, 176 (2006).

[7] G. Vignale, J. Supercond. Nov. Magn. **23**, 3 (2010).

[8] G. E. W. Bauer, A. H. MacDonald, and S. Maekawa, Solid State Comm. **150**, 459 (2010) and other articles in the same issue.

[9] W. L. Lee, S. Watauchi, V. L. Miller, R. J. Cava, and N. P. Ong, Phys. Rev. Lett. **93**, 226601 (2004).

[10] N. Hanasaki, K. Sano, Y. Onose, T. Ohtsuka, S. Iguchi, I. Kezsmarki, S. Miyasaka, S. Onoda, N. Nagaosa, and Y. Tokura, Phys. Rev. Lett. **100**, 106601 (2008).

[11] T. Miyasato, N. Abe, T. Fujii, A. Asamitsu, S. Onoda, Y. Onose, N. Nagaosa, Y. Tokura, Phys. Rev. Lett. **99**, 086602 (2007).

[12] Y. Pu, D. Chiba, F. Matsukura, H. Ohno, and J. Shi, Phys. Rev. Lett. **101**, 117208 (2008).

[13] M. Mizuguchi, S. Ohata, K. I. Uchida, E. Saitoh, and K. Takanashi, Appl. Phys. Express **5**, 093002 (2012).

[14] M. Weiler *et al*., Phys. Rev. Lett. **108**, 106602 (2012).

[15] J. S. Park, H. I. Yim, J. Y. Hwang, S. B. Lee, and T. W. Kim, J. Korean. Phys. Soc. **57**, 1672 (2010)

[16] S. Jeong and H. I. Yim, J. Korean Phys. Soc. **60**, 450 (2012).

[17] J. Y. Hwang, S. S. Kim, and J. R. Rhee, J. Magn. Magn. Mater. **310**, 1943 (2007).

[18] E. H. M. van der Heijden, K. J. Lee, Y. H. Choi, T. W. Kim, H. J. M. Swagten, C.-Y. You, and M. H. Jung, Appl. Phys. Lett. **102**, 102410 (2013).

# Figure Captions

**Fig. 1.** (a) Schematic setup for the Nernst effect measurement with a local heater. (b) Optical microscope image of the sample which shows the Hall bar and the heater. The width of the Hall bar is 10 μm and the distance between the adjacent voltage contacts is 20 μm. (c) and (d) $V_N$ and $V_H$ as a function of $B$ for $N=7$ and for $N=15$, respectively.

**Fig. 2.** (a) $V_N$ as a function of $B$ for different values of $I_h$. (b) The saturated value of $V_N$ as a function of $I_h$. (c) The saturated value of $V_N$ as a function of heater power in a log-log scale. The solid line is a best linear fit with a slope of 1.15.

**Fig. 3.** The coercive field $H_c$ as a function of $I_h$. The dashed lines are to guide the eyes. $H_c$ determined from the Hall measurements with a drive current of 50 μA is about 220G and 150G for $N=7$ and 15, respectively.

**Fig. 4.** (a) $V_N$ as a function of $B$ measured from different probing contacts with $I_h=60$ mA. $d$ is the distance from the heater to each pair of probing contacts. (b) The saturated value of $V_N$ as a function of $d$ in a linear scale and (c) in a log-log scale. The best linear fit in (c) represents $V_N \propto 1/d^{1.3}$.

**Fig. 1.**

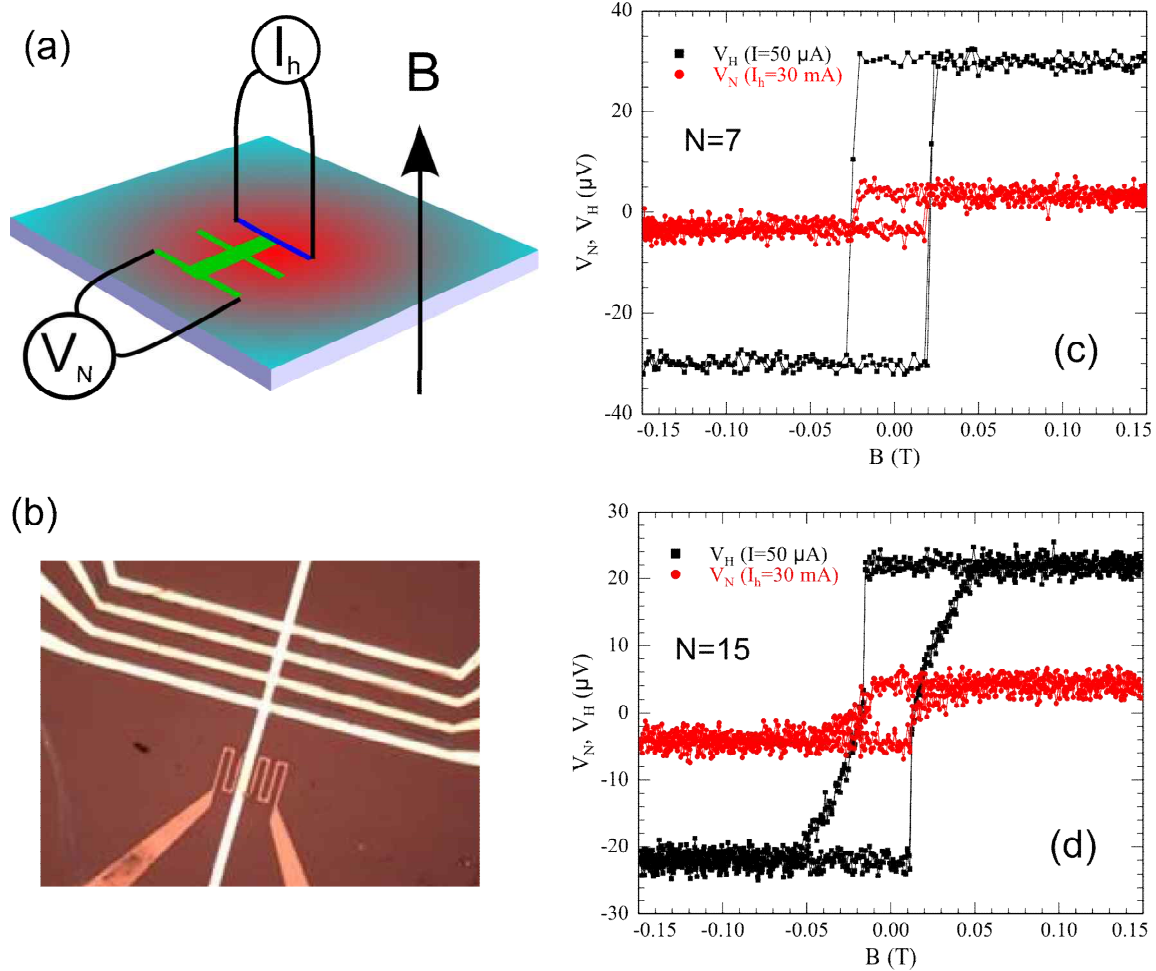

**Fig. 2.**

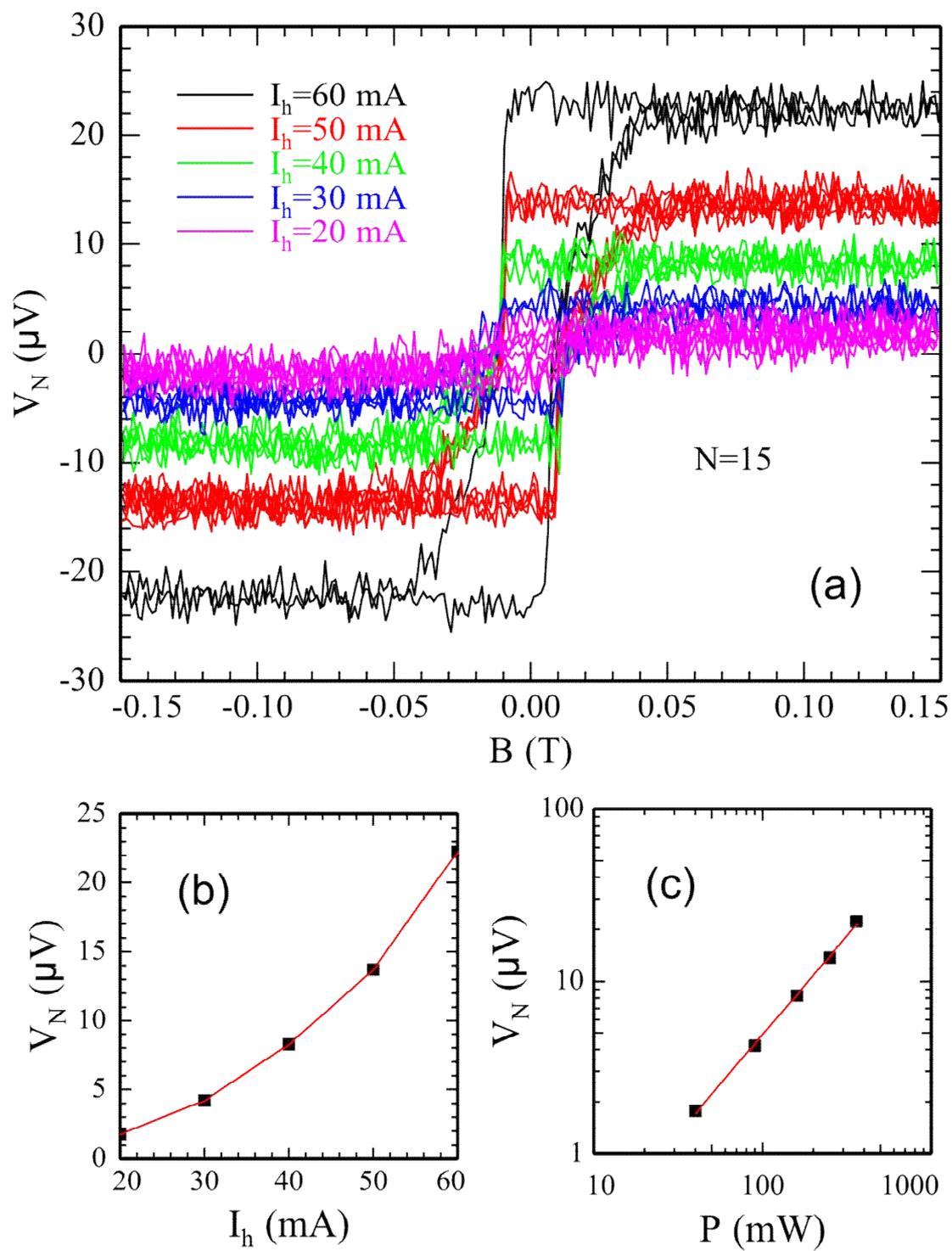

**Fig. 3.**

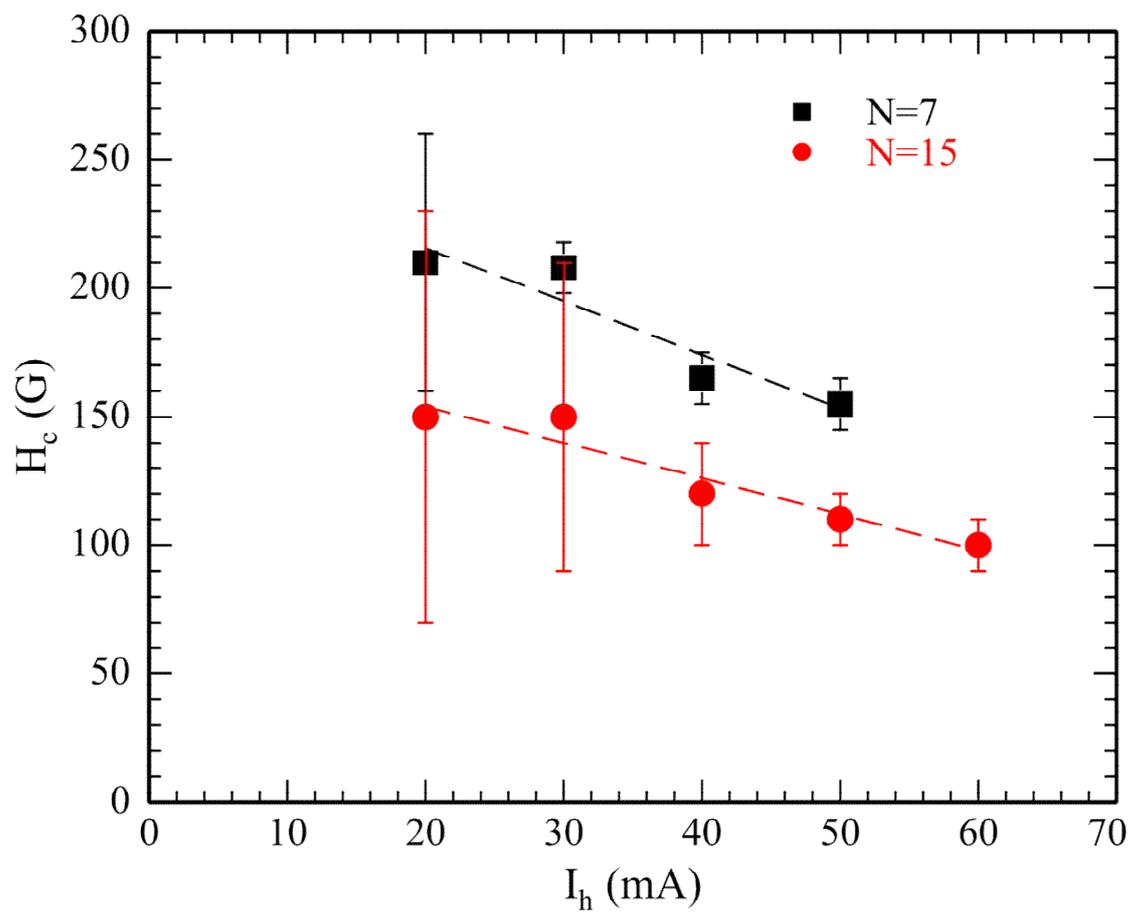

**Fig. 4.**

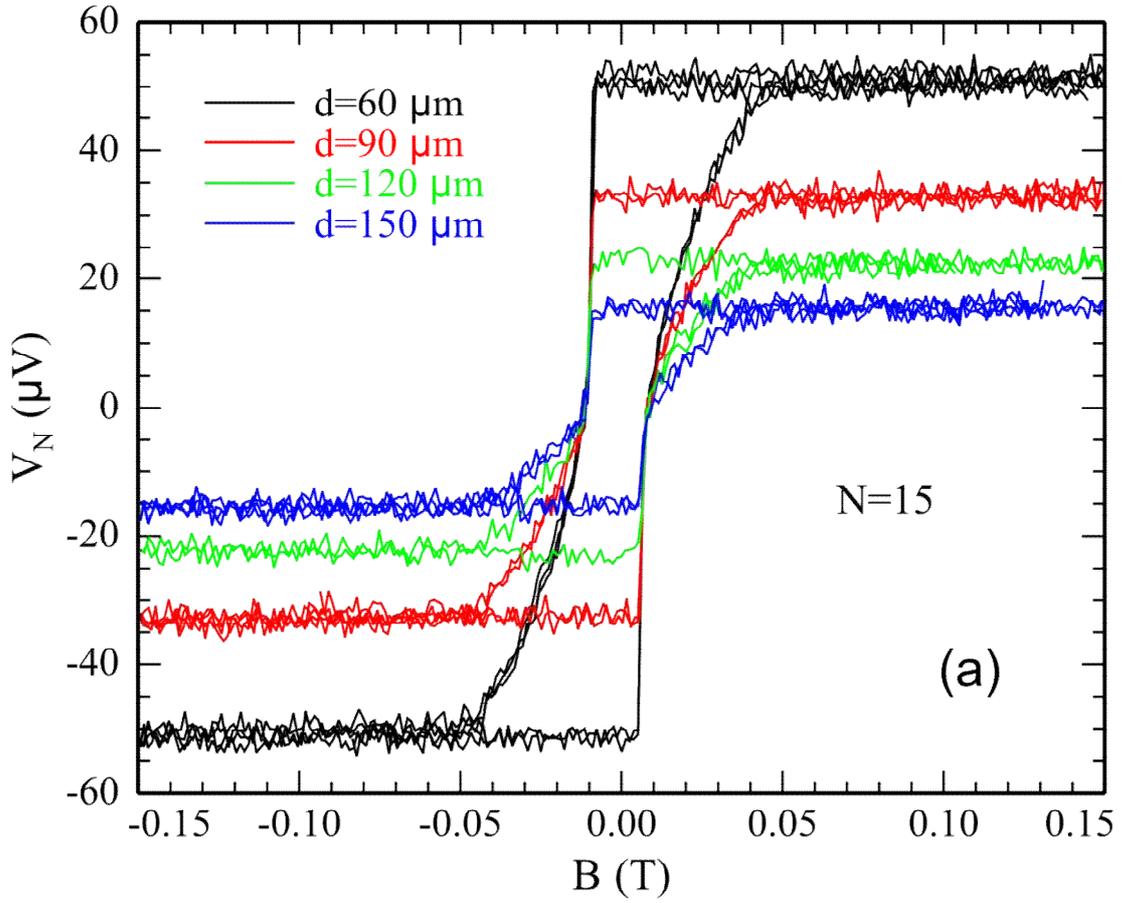

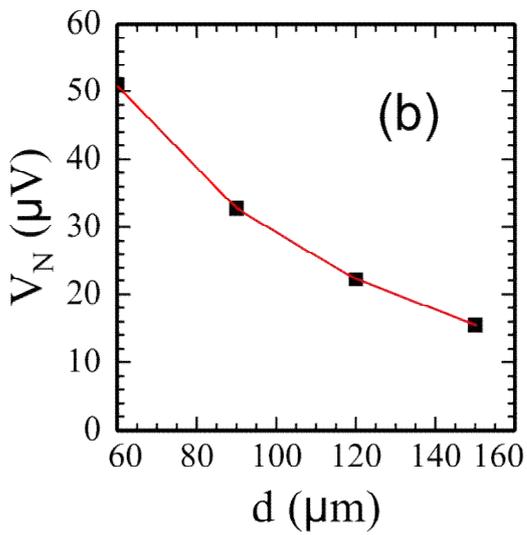 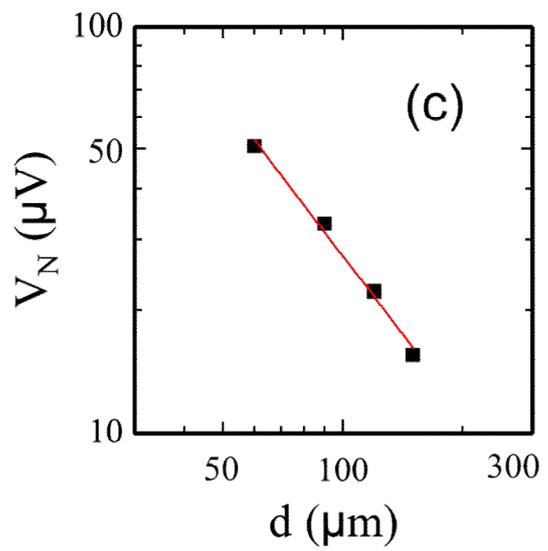